\documentclass[11pt, reqno]{amsart}
\usepackage{graphicx,amsmath,amssymb,epsfig}
\usepackage{amsfonts}
\usepackage{booktabs}
\usepackage{subcaption}

\usepackage{hyperref}

\usepackage[
backend=biber,
style=apa,
sorting=nyt
]{biblatex}
\DeclareSourcemap{
  \maps[datatype=bibtex]{
    \map{
      \step[fieldset=issn, null]
      \step[fieldset=doi, null]
      \step[fieldset=url, null]
      \step[fieldset=urldate, null]
    }
  }
}

\hypersetup{
    colorlinks=true,
    urlcolor=blue,
    linkcolor=blue,
    citecolor=blue
}

\newtheorem{theorem}{Theorem}
\theoremstyle{plain}

\theoremstyle{definition}

\numberwithin{equation}{section}
\graphicspath{ {FigsMKQR/} }

\DeclareMathOperator*{\argmax}{arg\,max}

\begin{document}
\title[Applications of Optimal Transport Methods to Econometrics]{A Survey
of Some Recent Applications of Optimal Transport Methods to Econometrics}
\author{Alfred Galichon}
\address{New York University, Economics Department, Faculty of Arts and
Sciences, and Mathematics Department, Courant Institute of Mathematical
Sciences}
\date{September, 2016. Galichon's research has received funding from the
European Research Council under the European Union's Seventh Framework
Programme (FP7/2007-2013) / ERC grant agreement 313699.}

\begin{abstract}
This paper surveys recent applications of methods from the theory of optimal
transport to econometric problems.
\end{abstract}

\maketitle


{\footnotesize \textbf{JEL Classification}: C01, C02.\vskip50pt }



\section{Introduction}

Optimal transport, popularized by Villani's texts~\cite{Villani03} and~\cite%
{Villani09}, is currently a very active research area of mathematics, and it
has found applications in many sciences. Economics is no exception. However,
up to a recent period, the appearance of optimal transport in economics was
only in connection with two-sided models of matching (see e.g. \cite{Becker}%
, \cite{shapleyShubik}, \cite{GOZ}): indeed, as shown by Chiappori et al.
\cite{CMN}, equilibrium outcomes in two-sided matching problems with
transferable utility coincide with the solutions of an optimal transport
problem. More recently however, methods from optimal transport theory have
been used as a tool in a number of problems in econometrics, microeconomic
theory, and finance. These methods are exposed in a comprehensive way in my
recent monograph, \emph{Optimal Transport Methods in Economics}~\cite%
{GalichonOTME}, aimed at an audience of economists. The goal of the present
paper, which partly follows the presentation there, is to provide a short
introduction to the use of optimal transport methods in econometrics. The
reader will hopefully forgive me for a bias toward my own research in
selecting applications.

\bigskip

The paper is organized as follows. Section 2 provides a brief overview
of the theory, and states its main result, the Monge-Kantorovich theorem.
Section 3 discusses three particular cases where, because of additional
restrictions, the analysis of solutions of the Monge-Kantorovich problem can
be pushed further, which is most helpful in applications. Section 4 reviews
a number of econometric applications involving optimal transport methods.

\section{Monge-Kantorovich theory in a nutshell}

\subsection{Optimal coupling}

We start by describing the optimal transport problem at an intermediate
level of generality, which will suffice for our applications. Let $\mathcal{X%
}$ and $\mathcal{Y}$ be two closed subsets of $\mathbb{R}^{d}$ and $\mathbb{R%
}^{d^{\prime }}$, respectively, and consider two (Borel) probability
distributions $P$ and $Q$ of respective supports $\mathcal{X}$ and $\mathcal{%
Y}$.  A \emph{coupling} of probabilities $P$ and $Q$ is a joint probability
distribution $\pi $ on $\mathcal{X}\times \mathcal{Y}$ with marginal
distributions $P$ and $Q$, which means that if $\left( X,Y\right) $ is a
random vector with probability distribution $\pi $, then its projections $X$
and $Y$ on $\mathcal{X}$ and $\mathcal{Y}$ should be random vectors with
respective probability distributions $P$ and $Q$. The set of such couplings
will be denoted as%
\begin{equation*}
\mathcal{M}\left( P,Q\right) =\left\{ \pi :\left( X,Y\right) \sim \pi \text{
implies }X\sim P\text{ and }Y\sim Q\right\} .
\end{equation*}

Hence, $\pi \in \mathcal{M}\left( P,Q\right) $ encodes the \textquotedblleft
missing information\textquotedblright\ needed in order to build a random
vector $\left( X,Y\right) $ when $X\sim P$ and $Y\sim Q$. When $\mathcal{X}=%
\mathcal{Y}=\left[ 0,1\right] $, and $P=Q=\mathcal{U}\left( \left[ 0,1\right]
\right) $, $\mathcal{M}\left( P,Q\right) $ coincides with the set of \emph{%
copulas}, which are well-known objects in applied probability whose purpose
is also to build bivariate distributions based on univariate ones.
Therefore, couplings can be seen as a generalization of copulas beyond the
univariate case.

\bigskip

Following Monge~\cite{monge} and Kantorovich \cite{kantorovich39,
kantorovich48}, we shall consider the \emph{Monge-Kantorovich problem} of
finding the \textquotedblleft optimal\textquotedblright\ coupling of
probability distributions $P$ and $Q$. By optimal, we mean here that it
should maximize the expectation of some surplus function $\Phi :\mathcal{X}%
\times \mathcal{Y}\rightarrow \mathbb{R}$, that is,
\begin{equation}
\sup_{\pi \in \mathcal{M}\left( P,Q\right) }\mathbb{E}_{\pi }\left[ \Phi
\left( X,Y\right) \right] .  \label{MK-primal}
\end{equation}

We can interpret this problem as a problem of worker-firm assignment: a
central planner needs to assign on a one-to-one basis a population of
workers, whose skills are distributed on set $\mathcal{X}$ with distribution
$P$, to a population of firms whose characteristics are distributed
according to $Q$ on set $\mathcal{Y}$. The economic value created by worker $%
x$ if employed by firm $y$ is $\Phi \left( x,y\right) $. An assignment of
workers to firms is defined as $\pi \in \mathcal{M}\left( P,Q\right) $,
which measures the distribution of the matches. The total economic value
created by an assignment $\pi $ is therefore $\mathbb{E}_{\pi }\left[ \Phi
\left( X,Y\right) \right] $, and hence, the maximum economic value that the
central planner can hope to achieve is~(\ref{MK-primal}).

\subsection{Duality and the Monge-Kantorovich theorem}

Problem~(\ref{MK-primal}) is an infinite-dimensional linear programming
problem. Indeed, the objective function is linear because it is an
expectation with respect to measure $\pi $, which is the optimization
variable, and the constraint $\pi \in \mathcal{M}\left( P,Q\right) $ can be
expressed as $\mathbb{E}_{\pi }\left[ \varphi \left( X\right) +\psi \left(
Y\right) \right] =\mathbb{E}_{P}\left[ \varphi \left( X\right) \right] +%
\mathbb{E}_{Q}\left[ \psi \left( Y\right) \right] $ for all integrable test
functions $\varphi :\mathcal{X}\rightarrow \mathbb{R}$ and $\psi :\mathcal{Y}%
\rightarrow \mathbb{R}$, and these constraints are linear with respect to $%
\pi $.

As a result, problem~(\ref{MK-primal}) has a dual formulation. The dual
formulation can be worked out by hand, by noticing that if $\varphi $ and $%
\psi $ are test functions such that $\Phi \left( x,y\right) \leq \varphi
\left( x\right) +\psi \left( y\right) $ for all $x$ and $y$ in their
domains, then by integration with respect to any $\pi \in \mathcal{M}\left(
P,Q\right) $ and by using the fact that $\pi $ has margins $P$ and $Q$, it
follows that
\begin{equation*}
\mathbb{E}_{\pi }\left[ \Phi \left( X,Y\right) \right] \leq \mathbb{E}_{P}%
\left[ \varphi \left( X\right) \right] +\mathbb{E}_{Q}\left[ \psi \left(
Y\right) \right] .
\end{equation*}

This being true for any $\pi \in \mathcal{M}\left( P,Q\right) $ and any test
functions $\varphi $ and $\psi $ such that $\Phi \left( x,y\right) \leq
\varphi \left( x\right) +\psi \left( y\right) $, it follows by taking the
maximum on the left and the minimum on the right hand side that the value of
problem~(\ref{MK-primal}) is less or equal to the value of the \emph{dual
problem}%
\begin{equation}
\inf_{\varphi \left( x\right) +\psi \left( y\right) \geq \Phi \left(
x,y\right) }\mathbb{E}_{P}\left[ \varphi \left( X\right) \right] +\mathbb{E}%
_{Q}\left[ \psi \left( Y\right) \right]  \label{MK-dual}
\end{equation}%
where the minimum is over all $P$- and $Q$- integrable functions $\varphi $
and $\psi $ such that $\varphi \left( x\right) +\psi \left( y\right) \geq
\Phi \left( x,y\right) $. Under a few additional assumptions on $P$, $Q$ and
$\Phi $, the Monge-Kantorovich theorem asserts that this inequality actually
holds as an equality, and the optimal coupling $\pi $ exists. Under an
additional assumption, pairs of solution potentials $\left( \varphi ,\psi
\right) $ to the dual problem~(\ref{MK-dual}) also exist. This is made
precise in the following result:

\begin{theorem}[Monge-Kantorovich duality theorem]
Assume $\Phi :\mathcal{X}\times \mathcal{Y}\rightarrow \mathbb{R}\cup
\left\{ -\infty \right\} $ is an upper semicontinuous surplus function
bounded from above by $\overline{a}\left( x\right) +\overline{b}\left(
y\right) $ where $\overline{a}$ and $\overline{b}$ are respectively
integrable with respect to $P$ and $Q$. Then:

(i) Strong duality holds, that is:%
\begin{equation}
\sup_{\pi \in \mathcal{M}\left( P,Q\right) }\mathbb{E}_{\pi }\left[ \Phi
\left( X,Y\right) \right] =\inf_{\varphi \left( x\right) +\psi \left(
y\right) \geq \Phi \left( x,y\right) }\mathbb{E}_{P}\left[ \varphi \left(
X\right) \right] +\mathbb{E}_{Q}\left[ \psi \left( Y\right) \right] ,
\label{MKDuality}
\end{equation}%
where the infimum on the right hand-side is taken over measurable and
integrable functions $\varphi $ and $\psi $, and the inequality constraint
should be satisfied for $P-$almost every $x$ and $Q-$almost every $y$.

(ii) An optimal solution $\pi $ to the primal problem on the left hand-side
exists.

(iii) Assume further $\Phi $ is bounded from below by $\underline{a}\left(
x\right) +\underline{b}\left( y\right) $ where $\underline{a}$ and $%
\underline{b}$ are respectively integrable with respect to $P$ and $Q$. Then
the dual problem on the right hand-side also has solutions.
\end{theorem}

For a proof of this result, refer to~\cite{Villani09}, Theorem 5.10.

\bigskip

Let us now explore a criterion for jointly checking that a coupling $\pi $
and a pair of potential functions $\left( \varphi ,\psi \right) $ are
simultaneously optimal for the primal and the dual respectively. Take $\pi
\in \mathcal{M}\left( P,Q\right) $ and take $\varphi $ and $\psi $ such that
$\varphi \left( x\right) +\psi \left( y\right) \geq \Phi \left( x,y\right) $%
. Then $\pi $ and $\left( \varphi ,\psi \right) $ are respectively solutions
to~(\ref{MK-primal}) and~(\ref{MK-dual}) if and only if the equality $%
\mathbb{E}_{\pi }\left[ \varphi \left( X\right) +\psi \left( Y\right) \right]
=\mathbb{E}_{\pi }\left[ \Phi \left( X,Y\right) \right] $ holds. This is
equivalent to the fact that
\begin{equation}
Supp\left( \pi \right) \subseteq \left\{ \left( x,y\right) \in \mathcal{X}%
\times \mathcal{Y}:\varphi \left( x\right) +\psi \left( y\right) =\Phi
\left( x,y\right) \right\} .  \label{supportCondition}
\end{equation}

In the worker-firm interpretation alluded to above, this condition has an
interpretation in terms of \emph{pairwise stability}. Indeed, it implies
that $\varphi \left( x\right) $ can be interpreted as the payoff of a worker
of type $x$, and $\psi \left( y\right) $ can be interpreted as the payoff of
a firm of type $y$. If $x$ and $y$ are matched at equilibrium, then $\left(
x,y\right) \in Supp\left( \pi \right) $, thus $\varphi \left( x\right) +\psi
\left( y\right) =\Phi \left( x,y\right) $, that is, the output $\Phi \left(
x,y\right) $ created by the $\left( x,y\right) $ pair should be divided into
the payoff of the worker $\varphi \left( x\right) $, and the payoff of the
firm $\psi \left( y\right) $. If $x$ and $y$ are not matched, then $\varphi
\left( x\right) +\psi \left( y\right) \geq \Phi \left( x,y\right) $, which
expresses the fact that $x$ and $y$ do not have an incentive to leave their
existing partners to form a blocking pair. Thus $\left( \varphi ,\psi
\right) $ is a pair of \textquotedblleft stable\textquotedblright\ payoffs.

\subsection{Some remarks}

Let us note a few immediate properties of this problem.

\bigskip

First, note that if one replaces $\Phi \left( x,y\right) $ by $\Phi \left(
x,y\right) +a\left( x\right) +b\left( y\right) $, then the set of optimal
couplings $\pi $ for problem~(\ref{MK-primal}) remains unchanged: indeed,
the value of $\mathbb{E}_{\pi }\left[ a\left( X\right) +b\left( Y\right) %
\right] $ does not depend on the choice of $\pi \in \mathcal{M}\left(
P,Q\right) $. Also note that if $\left( \varphi ,\psi \right) $ is a
solution to problem~(\ref{MK-dual}), then for any real number $c$, $\left(
\varphi -c,\psi +c\right) $ is also a solution.

\bigskip

Next, note that if a minimizer $\left( \varphi ,\psi \right) $ of the dual
problem~(\ref{MK-dual}) exists, then one has necessarily
\begin{eqnarray}
\varphi \left( x\right) &=&\max_{y\in \mathcal{Y}}\left\{ \Phi \left(
x,y\right) -\psi \left( y\right) \right\}  \label{phi} \\
\psi \left( y\right) &=&\max_{x\in \mathcal{X}}\left\{ \Phi \left(
x,y\right) -\varphi \left( x\right) \right\} ,  \label{psi}
\end{eqnarray}%
which has another interpretation, this time in terms of Walrasian
equilibrium: $\varphi \left( x\right) $ can be seen as the equilibrium wage
of a worker of type $x$, while $\psi \left( y\right) $ can be interpreted as
the equilibrium surplus of a firm of type $y$. In particular, formula~(\ref%
{psi}) expresses the problem of a firm $y\in \mathcal{Y}$ looking for a
worker $x\in \mathcal{X}$ yielding the highest surplus $\Phi \left(
x,y\right) -\varphi \left( x\right) $.

Formulas~(\ref{phi}) and~(\ref{psi}) allow to rewrite problem~(\ref{MK-dual}%
) using a number of alternative reformulations:

\begin{itemize}
\item Using the expression of $\varphi $ given in~(\ref{phi}), problem~(\ref%
{MK-dual}) rewrites as
\begin{equation}
\inf_{\psi }\left\{ \mathbb{E}_{P}\left[ \max_{y\in \mathcal{Y}}\left\{ \Phi
\left( X,y\right) -\psi \left( y\right) \right\} \right] +\mathbb{E}_{Q}%
\left[ \psi \left( Y\right) \right] \right\}  \label{usingPsiOnly}
\end{equation}

\item Using the fact that a constant can be freely added to or subtracted
from $\psi $, one may look at the solutions of problem~(\ref{MK-dual}) among
those such that $\mathbb{E}_{Q}\left[ \psi \left( Y\right) \right] =0$, that
is, problem~(\ref{MK-dual}) yet rewrites as%
\begin{eqnarray*}
\inf_{\psi } &&\mathbb{E}_{P}\left[ \max_{y\in \mathcal{Y}}\left\{ \Phi
\left( X,y\right) -\psi \left( y\right) \right\} \right] \\
s.t.~ &&\mathbb{E}_{Q}\left[ \psi \left( Y\right) \right] =0.
\end{eqnarray*}
\end{itemize}

These alternative formulations will turn out to be useful below.

\section{Three cases of interest}

\subsection{Discrete optimal assignment problem}

One first case of interest is found when $P$ and $Q$ are discrete
distributions with finite support: $P=\sum_{i=1}^{n}p_{i}\delta _{x_{i}}$
and $Q=\sum_{j=1}^{m}q_{j}\delta _{y_{j}}$. In which case, we denote $\Phi
_{ij}=\Phi \left( x_{i},y_{j}\right) $, and duality~(\ref{MKDuality})
rewrites as%
\begin{equation}
\begin{tabular}{lll}
$\max_{\pi \geq 0}\sum_{ij}\pi _{ij}\Phi _{ij}$ & $=$ & $\min_{\varphi ,\psi
}\sum_{i}p_{i}\varphi _{i}+\sum_{j}q_{j}\psi _{j}$ \\
$s.t.~\left.
\begin{array}{c}
\sum_{i}\pi _{ij}=p_{i} \\
\sum_{j}\pi _{ij}=q_{j}%
\end{array}%
\right. $ &  & $s.t.~\left. \varphi _{i}+\psi _{j}\geq \Phi _{ij}\right. $%
\end{tabular}
\label{LP}
\end{equation}

The complementary slackness condition states that if $\pi _{ij}$, which is
the Lagrange multiplier associated to the dual constraint $\varphi _{i}+\psi
_{j}\geq \Phi _{ij}$, is strictly positive, then the corresponding
constraint is saturated. Thus $\pi _{ij}>0$ implies $\varphi _{i}+\psi
_{j}\geq \Phi _{ij}$, which recovers the general condition~(\ref%
{supportCondition}) viewed above.

Formulated as a linear programming problem as in~(\ref{LP}), the optimal
transport problem can be solved using standard linear programming toolboxes;
see~\cite{GalichonOTME}, section 3.4 how to perform computations efficiently
using the sparse structure of the constraint matrix.

\subsection{Continuous-to-discrete case}

One second case of interest arises when $P$ is a continuous distribution,
but when $Q$ is a discrete distribution with finite support: $%
Q=\sum_{j=1}^{m}q_{j}\delta _{y_{j}}$. In this case, the reformulation of
the Monge-Kantorovich problem using expression~(\ref{usingPsiOnly}) is
particularly useful. Indeed, following Aurenhammer~\cite{aurenhammer87}, one
may then expresses problem~(\ref{MK-dual}) as%
\begin{equation}
\min_{\psi \in \mathbb{R}^{m}}\mathbb{E}_{P}\left[ \max_{j\in \left\{
1,...,m\right\} }\left\{ \Phi \left( X,y_{j}\right) -\psi _{j}\right\} %
\right] +\sum_{j=1}^{m}q_{j}\psi _{j},  \label{MKDiscreteContinuous}
\end{equation}%
which is simply the problem of minimizing a convex function in $\mathbb{R}%
^{m}$. See~\cite{GalichonOTME}, section 5.3 for a discussion on the
implementation of this method using gradient descent.

\subsection{Scalar product surplus\label{par:Brenier}}

A third case of interest is the case when $P$ and $Q$ are continuous
distributions on $\mathcal{X}=\mathcal{Y}=\mathbb{R}^{d}$ ($d=d^{\prime }$),
and when $\Phi \left( x,y\right) =x^{\intercal }y$ is the scalar product
between $x$ and $y$. In this case, under the assumption that $P$ and $Q$
have finite second moments, a solution $\left( \varphi ,\psi \right) $ to~(%
\ref{MK-dual}) exists, and relations~(\ref{phi}) and~(\ref{psi}) become%
\begin{equation*}
\left\{
\begin{array}{c}
\varphi \left( x\right) =\max_{y\in \mathbb{R}^{d}}\left\{ x^{\intercal
}y-\psi \left( y\right) \right\} \\
\psi \left( y\right) =\max_{x\in \mathbb{R}^{d}}\left\{ x^{\intercal
}y-\varphi \left( x\right) \right\}%
\end{array}%
\right.
\end{equation*}%
hence $\varphi $ and $\psi $ are related by Legendre-Fenchel conjugation,
which is classically denoted $\psi =\varphi ^{\ast }$ and $\varphi =\psi
^{\ast }$. A short tutorial on convex analysis from the point of view of
optimal transport is provided in~\cite{GalichonOTME}, section 6.1, where the
basic notions used in the sequel, such as Legendre transform and
subdifferential, are recapitulated.

By~(\ref{supportCondition}), if $\pi \in \mathcal{M}\left( P,Q\right) $ is
an optimal coupling and if $\left( \varphi ,\psi \right) $ is a dual
solution, then $\pi $ and $\left( \varphi ,\psi \right) $ are both optimal
if and only if the support of $\pi $ is included in the set of $\left(
x,y\right) $ such that $\varphi \left( x\right) +\varphi ^{\ast }\left(
y\right) =x^{\intercal }y$. But this relation is equivalently reexpressed in
convex analysis by the fact that $y\in \partial \varphi \left( x\right) $,
where $\partial \varphi $ denotes the subgradient of $\varphi $ at $x$.
Hence, $\pi $ and $\left( \varphi ,\psi \right) $ are both optimal if and only if%
\begin{equation*}
Supp\left( \pi \right) \subseteq \left\{ \left( x,y\right) \in \mathbb{R}%
^{d}\times \mathbb{R}^{d}:y\in \partial \varphi \left( x\right) \right\} .
\end{equation*}

However, it is a well-known fact in convex analysis\footnote{%
It follows from Rademacher's theorem; see \cite{Villani09}, theorem 10.8.}
that if $\varphi $ is convex, then the set of points where $\varphi $ is not
differentiable is of zero Lebesgue measure. As a result, if $P$ is
absolutely continuous with respect to the Lebesgue measure, then $\partial
\varphi \left( x\right) $ $P$-almost surely coincides with $\left\{ \nabla
\varphi \left( x\right) \right\} $. Therefore, in this case, an optimal
coupling $\left( X,Y\right) \sim \pi $ can be represented by $\left(
X,\nabla \varphi \left( X\right) \right) $, where $\varphi $ is convex and
is such that $\nabla \varphi \left( X\right) \sim Q$. This is denoted as
\begin{equation}
\nabla \varphi \#P=Q  \label{pushfwd}
\end{equation}%
and one says that $\nabla \varphi $ \textquotedblleft pushes forward $P$
onto $Q$.\textquotedblright\ The existence and uniqueness (up to a constant)
of a convex function $\varphi $ satisfying~(\ref{pushfwd}) are the object of
Brenier's theorem \cite{Brenier87} (see also \cite{knottsmith} and \cite%
{RuschendorfRachev90}). This theorem was improved by McCann (\cite{mcCann95}%
), who obtained the existence of $\varphi $ in~(\ref{pushfwd}) without
requiring $P$ and $Q$ to have second moments.

\section{Econometric applications}

\subsection{Demand inversion in discrete choice models}

Our first application is the problem of demand inversion in discrete choice,
or random utility models. Consider a discrete choice model, where agents
face alternatives $y\in \mathcal{Y}$. The systematic utility associated to
alternative $y$ is a real number $\delta _{y}$, and the unobserved
heterogeneity associated to it is $\varepsilon _{y}$. It is assumed that $%
\varepsilon \sim \mathbf{P}$, where $\mathbf{P}$ is a probability
distribution on $\mathbb{R}^{\mathcal{Y}}$.

Let $q$ be a probability vector on $\mathcal{Y}$. One says that $q$ is a
vector of market shares induced by systematic utility vector $\delta $ if $q$
is the probability distribution of a random variable $Y$ such that $Y\in
\arg \max_{y\in \mathcal{Y}}\left\{ \delta _{y}+\varepsilon _{y}\right\} $.
The problem of demand inversion in discrete choice models, as defined for
instance in~\cite{Berry94},\ is as follows: given a vector of market shares $%
q$, what is the set of systematic utility vectors $D\left( q\right) $ whose
elements $\delta $ induce the choice probability $q$. Formally,%
\begin{equation*}
D\left( q\right) =\left\{ \delta \in \mathbb{R}^{\mathcal{Y}}:\exists Y\sim
q,Y\in \arg \max_{y\in \mathcal{Y}}\left\{ \delta _{y}+\varepsilon
_{y}\right\} \right\} .
\end{equation*}

\bigskip

The main observation here, due to~\cite{GalichonSalanie}, is that $\delta
\in D\left( q\right) $ if and only if $\psi =-\delta $ appears in the
solution of the dual Monge-Kantorovich problem~(\ref{MK-dual}). Indeed,
define%
\begin{equation*}
G\left( \delta \right) =\mathbb{E}_{\mathbf{P}}\left[ \max_{y\in \mathcal{Y}%
}\left\{ \delta _{y}+\varepsilon _{y}\right\} \right]
\end{equation*}%
then the envelope theorem shows that $\delta \in D\left( q\right) $ if and
only if $q\in \partial G\left( \delta \right) $. But according to a basic
result in convex analysis (e.g. \cite{GalichonOTME}, section 6.1), $q\in
\partial G\left( \delta \right) $ if and only if $\delta \in \partial
G^{\ast }\left( q\right) $; thus $D\left( q\right) =\partial G^{\ast }\left(
q\right) $, and
\begin{equation*}
D\left( q\right) =\arg \min_{\delta }\left\{ G\left( \delta \right)
-\sum_{y\in \mathcal{Y}}q_{y}\delta _{y}\right\}
\end{equation*}%
Thus, $\delta \in D\left( q\right) $ if and only if $\psi =-\delta $
minimizes $G\left( -\psi \right) +\sum_{y\in \mathcal{Y}}q_{y}\psi _{y}$,
that is, if and only if it solves%
\begin{equation*}
\min_{\psi }\mathbb{E}_{\mathbf{P}}\left[ \max_{y\in \mathcal{Y}}\left\{
\varepsilon _{y}-\psi _{y}\right\} \right] +\sum_{y\in \mathcal{Y}}q_{y}\psi
_{y}
\end{equation*}%
which is exactly problem~(\ref{MKDiscreteContinuous}) with $\Phi \left(
\varepsilon ,y\right) =\varepsilon _{y}$. Therefore the problem of demand
inversion in discrete choice models is equivalent to a optimal transport
problem, and numerical methods for solving the latter can be used for the
former. See the papers~\cite{GalichonSalanie},~\cite{ChiGalShu}, and an
overview in~\cite{GalichonOTME}, section 9.2.

\subsection{Multivariate quantiles}

Quantiles play a fundamental role in econometrics and applied statistics.
They are useful for comparing distributional outcomes, measuring risk and
inequality, identifying willingness-to-pay, etc. One of the most stringent
limitations of quantiles is the fact that they are fundamentally univariate
objects: indeed, the quantile function of real-valued random variable $Y$ is
defined as the inverse map of its cumulative distribution function, which
can only be inverted when $Y$ is univariate.

There is a considerable literature aiming at providing various
generalizations of the notion of quantile to the multivariate case, which we
will not review here. Our point here is to show that optimal transport
provides a sensible such generalization, see~\cite{GalichonOTME},
section~6.3.

One possible way to define the (univariate) quantile map is as the monotone
map which pushes forward the uniform distribution on $\left[ 0,1\right] $ to
a distribution of interest; in other words, the quantile map associated to
random variable $Y\sim \nu $ is the map $T$ such that (i) $T$ is
nondecreasing and such that if $U\sim \mathcal{U}\left( \left[ 0,1\right]
\right) $, then $T\left( U\right) $ has the same distribution $\nu $ as $Y$.
This point of view will be the basis of our multivariate generalization of
the notion of quantiles to the case of a random vector of dimension $d$:
letting $\mu $ be a probability distribution of reference over $\mathbb{R}%
^{d}$ (when $d=1$, a natural choice is the uniform distribution), the $\mu $%
\emph{-quantile map} associated to $\nu $ is the map $T:\mathbb{R}%
^{d}\rightarrow \mathbb{R}^{d}$ such that:

\begin{itemize}
\item $T\#\mu =\nu $, which means that $T$ pushes forward the distribution $%
\mu $ to $\nu $; and such that

\item $T$ is \textquotedblleft monotone\textquotedblright\ in the sense that
it is the gradient of a convex function: \thinspace $T=\nabla \varphi $,
where $\varphi :\mathbb{R}^{d}\rightarrow \mathbb{R}$.
\end{itemize}

The shape restriction that $T$ should be the gradient of a convex function
is a natural generalization as, in dimension one, monotone maps are the
gradients/derivatives of convex functions. That a unique solution to this
requirement exists and is unique is the object of McCann's theorem referred
above.

\bigskip

Let us now discuss how the $\mu $-quantile associated to an empirical
distribution is constructed. Let $\left\{ y_{1},...,y_{n}\right\} $ be a
sample, and let $\nu _{n}=n^{-1}\sum_{i=1}^{n}\delta _{y_{i}}$ be the
associated empirical distribution. Because the quantile map is constructed
as the gradient of a convex function, it is easy to see that in the case $%
\nu $ has a finite support, the quantile map should be the gradient of a
piecewise affine and convex function -- indeed, the gradient of such a map
will take a finite number of possible values. In this case, the quantile map
$T_{n}$ associated to $\nu _{n}$ will be defined as%
\begin{equation*}
T_{n}\left( u\right) =\argmax_{y_{i}\in \left\{
y_{1},...,y_{n}\right\} }\left\{ u^{\top }y_{i}-\psi _{i}\right\}
\end{equation*}%
where the weights $\psi _{i}$ form a solution to problem~(\ref%
{MKDiscreteContinuous}) with the surplus $\Phi $ chosen as the scalar
product. Hence, the empirical quantile map can be obtained as a
finite-dimensional optimization problem. Consistency of $T_{n}$, which
expresses that the $\mu $-quantile of $\nu _{n}$ converges to the $\mu $%
-quantile of $\nu $ has been shown in~\cite{CGHH}.

\bigskip

This definition of $\mu $-quantiles, which was originally introduced in~\cite%
{EGH}, has a number of applications, including:

\begin{itemize}
\item A\ notion of multivariate comonotonicity, which allows to construct
multivariate measures of financial exposures. This was the original
motivation in~\cite{EGH}, see also~\cite{BG}. The definition is that $Y_{1}$
and $Y_{2}$ are $\mu $-comonotone if the $\mu $-quantile of $Y_{1}+Y_{2}$ is
the sum of the $\mu $-quantiles of $Y_{1}$ and $Y_{2}$.

\item Multivariate counterparts for a number of stochastic orders, which are
known to rely on the notion of quantiles in the univariate case, such as
first-order stochastic dominance. See~\cite{CGH}.

\item An extension of the theory of rank-dependent expected utility to
multivariate risky prospects. Rank-dependent utility functions were built as
a response to paradoxes in expected utility theory such as Allais' paradox.
A multivariate extension of Yaari's utility function was proposed in~\cite%
{GHJET}.

\item A characterization of Pareto efficient risk-sharing arrangements in
the multivariate case, as was done in~\cite{CDG}, extending a result by \cite%
{LandsbergerMeilijson} of the efficient risk-sharing arrangements as the
comonotone ones.

\item An extension of Matzkin's quantile-based identification results in
hedonic models (see e.g. \cite{matzkin2003} and \cite{Heckman10}) to the
case with more than one attribute, as is done in~\cite{CGHP}.

\item A multivariate version of quantile regression based on a
semiparametric extension of the Monge-Kantorovich case proposed in~\cite{CCG}%
.
\end{itemize}

\subsection{Partial identification}

Optimal transport can also be a useful tool to handle partial identification
in incomplete models. Consider an economic model with parameter $\theta \in
\Theta $ which predicts that the population's income $X$ will have
distribution $P_{\theta }$. Assume that the income $X$ is not perfectly
observed, but that only the tax bracket in which the income belongs is
observed. If $y$ is the mid-point of the bracket, let $\Gamma \left(
y\right) =\left[ l\left( y\right) ,u\left( y\right) \right] $ be the
corresponding bracket. The brackets are indexed by their mid-point $Y$,
whose distribution $Y\sim Q$ is assumed to be observed. The identified set $%
\Theta _{I}$ is therefore the set of parameters $\theta \in \Theta $ such
that the distribution $X\sim P_{\theta }$ predicted by the model is
compatible with the observed distribution of the brackets $Y\in Q$. (We
abstract away from any sample uncertainty here). More precisely, $\Theta _{I}
$ is the set of $\theta $ such that there is a joint probability $\pi \in
\mathcal{M}\left( P_{\theta },Q\right) $ such that $\pi \left( X\in \Gamma
\left( Y\right) \right) =1$.

This compatibility problem can be formulated as an optimal transport
problem. Indeed, if
\begin{equation*}
\Phi \left( x,y\right) =1\left\{ x\in \Gamma \left( y\right) \right\}
\end{equation*}%
then $\theta \in \Theta _{I}$ if and only if
\begin{equation}
\max_{\pi \in \mathcal{M}\left( P_{\theta },Q\right) }\mathbb{E}_{\pi }\left[
1\left\{ X\in \Gamma \left( Y\right) \right\} \right] =1.
\label{compatibility}
\end{equation}

Problem~(\ref{compatibility}) is an optimal transport problem, and thus the
numerical determination of $\Theta _{I}$ boils down to the computation of
the value of such a problem. This equivalence has been put to use in~\cite%
{GHRestud} and~\cite{EGH-ET}; see a synthetic presentation in~\cite%
{GalichonOTME}, section 9.1.

\subsection{Revealed preference}

There is an interesting connection between optimal transport and Afriat's
theorem on revealed preference inequalities~\cite{afriat}. Indeed, the
problem of revealed preference under its most basic form consists in the
following: given the observation of $n$ bundles $x_{1},...,x_{n}$ in $%
\mathbb{R}^{d}$, and given corresponding price vectors $p_{1},...,p_{n}$ in $%
\mathbb{R}^{d}$, one would like to recover nontrivial utility functions $u$
such that
\begin{equation*}
x_{i}=\arg \max_{x_{k}:p_{i}^{\top }x_{j}\leq p_{i}^{\top }x_{i}}u\left(
x_{j}\right) .
\end{equation*}%
In the affirmative, one shall say that observations $\left\{ \left(
x_{i},p_{i}\right) \right\} $ are rationalizable. As shown by Afriat, an
affirmative answer to this problem is equivalent to the existence of $%
\lambda \in \mathbb{R}_{+}^{n}$, $\lambda \neq 0$ and $u\in \mathbb{R}^{n}$
such that
\begin{equation}
u_{i}-u_{j}\geq \lambda _{i}p_{i}^{\top }\left( x_{i}-x_{j}\right) .
\label{Afriat}
\end{equation}

As it was pointed out in~\cite{EG}, see also~\cite{KKN}, this problem can be
reformulated using an optimal transport problem. Let $p$ be the vector of
uniform probability over $\left\{ 1,...,n\right\} $, i.e. $p_{i}=1/n$ for $%
i=1,...,n$. Let $\Delta $ be the set of $\lambda \in \mathbb{R}_{+}^{n}$
such that $\sum_{i=1}^{n}\lambda _{i}=1$. For $\lambda \in \Delta $, set
\begin{equation}
W\left( \lambda \right) =\max_{\pi \in \mathcal{M}\left( p,p\right)
}\sum_{1\leq i,j\leq n}\pi _{ij}\Phi _{ij}^{\lambda }  \label{primalW}
\end{equation}%
where $\Phi _{ij}^{\lambda }=\lambda _{i}p_{i}^{\top }\left(
x_{i}-x_{j}\right) $. By Monge-Kantorovich duality, we have
\begin{equation*}
W\left( \lambda \right) =\frac{1}{n}\min \left\{
\sum_{i=1}^{n}u_{i}+\sum_{j=1}^{n}v_{j}:u_{i}+v_{j}\geq \Phi _{ij}^{\lambda
}\right\} .
\end{equation*}

\bigskip

Note that one has $W\left( \lambda \right) \geq 0$ as can be seen by taking $%
\pi _{ij}^{\ast }=1\left\{ i=j\right\} /n$ in expression~(\ref{primalW}).
Assume that $W\left( \lambda \right) =0$. Then it means that $\pi
_{ij}^{\ast }$ is optimal. Take a pair $\left( \phi ,\psi \right) $ which is
optimal for the dual problem. Thus by complementary slackness, $%
u_{i}+v_{i}=\Phi _{ii}^{\lambda }=0$, hence $u_{i}=-u_{i}$. Thus the dual
feasibility condition then implies Afriat's inequalities~(\ref{Afriat}).
Conversely, it is quite easy to show that if Afriat's inequalities are
satisfied, then $W\left( \lambda \right) =0$. This provides a particularly
simple criterion: observations $\left\{ \left( x_{i},p_{i}\right) \right\} $
are rationalizable if and only if%
\begin{equation*}
\min_{\lambda \in \Delta }W\left( \lambda \right) =0
\end{equation*}%
which is a convex minimization problem. Further, the subgradient of $W\left(
\lambda \right) $ is the set of $Z\left( \pi \right) \in \mathbb{R}^{d}$
such that $Z_{i}\left( \pi \right) =p_{i}^{\top }\left(
x_{i}/n-\sum_{j=1}^{n}\pi _{ij}x_{j}\right) $, for any $\pi $ solution of~(%
\ref{primalW}).

\section{Conclusion}

This short article has hopefully convinced the reader of the growing
importance of optimal transport methods as part of the standard
econometrician's toolbox. Because it is so intrinsically connected with
notions such as linear programming, convex analysis, duality, quantiles,
copulas, clustering, graphs, and numerical methods, investing some time in
the study of this theory can only be fruitful. While this article has kept a
focus on econometrics, a number of other economic applications of optimal
transport exist, notably in mechanism design, labor economics, family
economics, and asset pricing. Some of these applications are reviewed
in \emph{Optimal Transport Methods in Economics}~\cite{GalichonOTME}.

\newpage
\printbibliography

\end{document}